\documentstyle[aps,prb,psfig]{revtex}
\renewcommand{\vec}[1]{{\bf #1}}

\newcommand{\wsf}{\omega_{\rm sf}}

\newcommand{\cdw}{\chi_{\rm c}}
\newcommand{\sdw}{\chi_{\rm s}}
\newcommand{\YBCO}{\mbox{$\rm YBa_2Cu_3O_{7-\delta}$}}
\newcommand{\RT}{R_{\rm T}}

\newcommand{\be}{\begin{equation}}
\newcommand{\ee}{\end{equation}}
\newcommand{\bea}{\begin{eqnarray}}
\newcommand{\eea}{\end{eqnarray}}

\newcommand{\PRB}[3]{Phys. Rev. B {\bf #1}, #2 (#3)}
\newcommand{\PRL}[3]{Phys. Rev. Lett. {\bf #1}, #2 (#3)}
\newcommand{\ea}{\mbox{\em et al.}}

\begin{document}
\draft
\title{%
Magnetic Properties of ${\rm YBa_2Cu_3O_{7-\delta}}$
in a self-consistent
approach:
Comparison with  Quantum-Monte-Carlo Simulations and Experiments
}
\author{G.~Hildebrand, E.~Arrigoni, C.~Gr\"ober,  and W.~Hanke}
\address{Institut f\"ur Theoretische Physik, 
Universit\"at W\"urzburg, Am
Hubland, 97074 W\"urzburg, Germany}

\date{To appear in Phys. Rev. B (sched. Feb. 99)}

\maketitle
\begin{abstract}
We analyze single-particle electronic and 
two-particle magnetic properties of the Hubbard
model in the underdoped and optimally-doped regime of \YBCO by means
of a modified version of the
fluctuation-exchange approximation, which only includes particle-hole
fluctuations. 
Comparison of our results 
with Quantum-Monte Carlo (QMC) calculations at relatively high  
temperatures ($T\sim 1000 K$) suggests  to
 introduce   a temperature renormalization in order 
to  improve the agreement between the two methods
 at intermediate and large values of the interaction $U$.
 We evaluate the
temperature dependence of the 
spin-lattice relaxation time $T_1$  and of the spin-echo
decay time $T_{2G}$ 
and compare it with
the results of NMR measurements on an underdoped and an optimally
doped \YBCO sample. 
For $U/t=4.5$ it is possible to consistently 
adjust the parameters of the Hubbard model in order to have a good
 {\it semi-quantitative} 
description 
of this temperature dependence
for temperatures larger than the spin gap as obtained from NMR measurements.
We also discuss the case $U/t\sim 8$, which is more appropriate
to  describe magnetic and single-particle properties close to half-filling.
However, for this larger value of $U/t$ 
the agreement with QMC as well as with experiments at finite doping is less
satisfactory.

\end{abstract}
\pacs{PACS numbers: 
71.27.+a, %Strongly correlated electron systems; heavy fermions
74.72.-h  %High-T sub c compounds
76.60.-k %Nuclear magnetic resonance and relaxation
}

\widetext
\section{Introduction}
\label{intro}
Intensive research of the
last decade made clear that antiferromagnetism and
superconductivity are the two dominating properties of  high-temperature
superconductors. Indeed, 
the fact that these
two states of matter
do not exclude each other and that their fluctuations coexist in an
extended parameter range 
suggests 
a close relation between them.
This has been the main motivation for the recently proposed SO(5) theory of
superconductivity which unifies antiferromagnetism and superconductivity on
the basis of a common symmetry principle.\cite{zhan.97,ed.ha.97,me.ha.97}
Here, as well as in more phenomenological approaches to the high-$T_c$
compounds\cite{scal.physrep,MMP,PinesZfP,BP,ZBP,si.zh.93}
which relate their underdoped properties to remnants of the
antiferromagnetic order, the key to understanding the driving mechanism
behind superconducting pairing lies in their magnetic properties.
The minimal microscopic model considered 
to describe  strong correlations effects is the Hubbard
model with nearest-neighbor  hopping $t$ 
and  on-site repulsion $U$.
Using Quantum-Monte-Carlo (QMC) simulations combined with Maximum Entropy 
techniques\cite{pr.ha.95,du.na.97} this model has recently been shown to
reproduce salient features in the underdoped photoemission experiment in
particular the pseudogap and its doping and momentum
dependence.\cite{du.mo.95} The latter has uniquely been related to the
momentum and doping dependence of magnetic correlations.\cite{pr.ha.97}
In order to describe more accurately the Fermi surface of the cuprate
materials, which appears to be closed around the  antiferromagnetic 
point $(\pi,\pi)$,\cite{lspacing} as shown in
several high-resolution angular-resolved-photoemission
spectroscopy (ARPES) measurements~\cite{ShenReview}
on $\rm Sr_2CuO_2Cl_2$ \cite{ki.wh.98},
$\rm YBa_2Cu_3O_{7-\delta}$ (YBCO) with $\delta\approx 0.1$\cite{Liu} and 
$\rm Bi_2Sr_2CaCu_2O_{8+\delta}$ (Bi2212),\cite{Dessau,ma.de.96} it has been
suggested to extend the model by
introducing
additional longer-range hoppings, namely  second ($t'$) and
 third ($t''$) nearest neighbors.
%%@@
This has been enforced by comparison of ARPES data with numerical
calculations on the $t-J$ model 
\cite{to.ma.94,an.li.95,be.ch.96,xi.wh.96,ed.oh.97.d,ki.wh.98}.
Qualitatively, a large range of values of $t'$ and $t''$ yield a 
Fermi-surface with the appropriate shape. However, it is pertinent to 
specify more precisely the values of the parameters $t'$ and $t''$
which give simultaneously 
a good {\it qualitative} description of other properties, in
 particular magnetism. %\\
In the present work, we combine two different many-body techniques,
namely, QMC, and
a modified version of the FLEX approximation\cite{bi.sc.89} 
(here referred to as MFLEX, for clarity)
whereby  particle-particle fluctuations, as well as 
vertex corrections
in two-particle correlations\cite{bi.wh.91}
 are neglected. The two techniques 
have advantages and disadvantages in different parameter regimes (Coulomb
correlation $U/t$, temperature $T$, system size $N$). Here, we 
 intend to
provide a definite link between single-particle, i.e. photoemission (ARPES)
and two-particle, i.e. magnetic excitations. This link may be a useful
guide and serve as an input not only in the unifying SO(5) theory but also
in phenomenological constructs such as the nearly antiferromagnetic Fermi
liquid theory (NAFL).\cite{MMP,PinesZfP}
In practice,
we  carry out a diagrammatic, self-consistent study of  
 single- and two-particle response functions of the Hubbard model
with inclusion of  longer-range
hopping terms up to third neighbors $t'$ and $t''$.
Our aim is to find a reasonable set of parameters for the model,
which consistently describe
at the same time magnetic (NMR) and electronic (ARPES) properties of
doped $\YBCO$ compounds, in particular, the Fermi-surface, the band
dispersion,  the spin-lattice relaxation time $T_1$  and the spin-echo
decay time 
$T_{2G}$.
More specifically,
we want to adjust these parameters, in particular $t'$ and $t''$, such
that the magnetic properties are reproduced at least in a semi-quantitative
way (and not just qualitatively) at least within a reasonable error.
We will show indeed that a careful tuning of the parameters is important
since a change  of $t'$ and $t''$ by only $25\%$ changes the result for the
$T_1$ and $T_{2G}$ by $100\%$ or more.
Since we want to compare 
theoretical and experimental results at moderate 
antiferromagnetic correlation lengths, it is important to perform the numerical
calculations at large system sizes at such low temperatures, 
which at present are not 
accessible by Quantum Monte-Carlo calculations (at least for
{\it dynamical} correlation functions). 
For this reason, we will use a 
refined diagrammatic technique
(MFLEX), whereby particle-hole diagrams with self-consistently
determined Greens functions are summed 
(See, e.g., Ref. \onlinecite{la.sc.95}),
 which
allows us to work on  $64 \times64$ (and even larger)
lattices down to temperatures $T\sim t/50$. 
Since we want to carry out the calculation with values of the
interaction of the order of the system's bandwith, where perturbational
approaches are uncontrolled, it is important to compare our results
with Quantum-Monte-Carlo (QMC) calculations, which provide essentially
exact results, in the temperature range accessible to this method.
We will show that up to intermediate values of $U$ ($U/t\lesssim 6$)
our diagrammatic results agree quite well with QMC
provided one allows for a renormalization of  the temperature $T$.
Our idea thus amounts to use the MFLEX calculation 
to extrapolate QMC calculations to low temperatures and large system
sizes,
which are not reachable by QMC simulations. 

The  values of $t'$ and $t''$ giving the best agreement with NMR
results turn out to depend on $U/t$. Consistent results are obtained for
$U/t=4.5$, $t=250 {\rm meV}$, $t'/t=-0.2$, and $t''/t=0.15$. For $U/t=8$
we need a larger $|t'/t|$, namely $t'/t=-0.4$,
 although the comparison with experiments
is less satisfactory in this case,
possibly due to the fact that our diagrammatic
calculation is less reliable for large $U/t$.
%%@@
These values of $t'$ and $t''$, especially the ones obtained for
$U/t=8$ are quite similar to the ones obtained 
by several authors who compared  ARPES data with numerical
calculations on the $t-J$ model
\cite{to.ma.94,an.li.95,be.ch.96,xi.wh.96,ed.oh.97.d,ki.wh.98}.

Our
paper consists of two main parts.  
In the first part we compare the diagrammatic results with QMC, while
in the second part we describe the NMR experiments.
More specifically, 
in Sec.~\ref{model} we shortly introduce the Hubbard model and describe the
MFLEX approximation.
We then discuss single-particle properties like the Fermi-surface
and the quasiparticle dispersion in Sec.~\ref{single}, 
followed by a detailed comparison of 
MFLEX and QMC results including single- and  two-particle properties in
Sec.~\ref{flex_qmc}.
In Sec.~\ref{flex_exp}, we justify our choice of the hopping parameters by
comparing our numerical results with NMR data on $\YBCO$.
Finally, we summarize and draw our conclusions in Sec.~\ref{summary}.

\section{Model and Technique}
\label{model}

The Hamiltonian of the Hubbard model is given  by:
\be
H= \sum_{\vec{k}, \sigma} (\epsilon_{\vec{k}} - \mu )  \;
c^\dagger_{\vec{k}, \sigma}
c_{\vec{k}, \sigma}
+ U \sum_i n_{i, \uparrow}  n_{i, \downarrow} \; ,
\label{eq:Hubbard_Hamiltonian}
\ee
where the bare energy dispersion
\bea
\epsilon_{\vec{k}} &=& -2 t  \left( \cos k_x  + \cos k_y \right)
-4 t'  \cos k_x   \cos k_y\nonumber \\
&&-2 t'' \left[ \cos(2 k_x) + \cos(2 k_y) \right]
\label{eq:bare_dispersion}
\eea
includes nearest-neighbor ($t$) and longer-range hopping processes
($t'$,$t''$). Here, $c_{\vec k \sigma}$ ($c^{\dagger}_{\vec k \sigma}$)
annihilates (creates) an electron with momentum 
$\vec k$ and spin $\sigma$. The chemical potential $\mu$  adjusts
the mean particle number $\langle n \rangle$ with the doping $x$
so that $\langle n \rangle =1 -x$.

The  fluctuation exchange approximation 
includes the interaction of the electrons with density, spin, and pairing
fluctuations in infinite order. According to the conserving
approximation scheme in the  Baym and Kadanoff\cite{BaKa} sense,
the self-energy
is obtained by differentiating an approximate  generating functional 
with respect to
the full Green's function. 
For the approximation to be  conserving in the two-particle channel it
is also necessary  to
 calculate the two-particle interaction by taking the
second functional derivative of the same generating functional with
respect to the Green's function. 
This leads to a rather complicated set of coupled equations which
can be solved only on small systems.~\cite{bi.wh.91}
Within the MFLEX approximation,
the  electron self-energy $\Sigma$ evaluated in the imaginary
(Matsubara) frequency representation reads\cite{bi.sc.89}:
\be
\label{sigma}
\Sigma(\vec{k},i \omega_m) = \frac{T}{N} \sum_{\vec{q}, \nu_n}
V(\vec{q},i \nu_n) G(\vec{k}-\vec{q},i \omega_m -i \nu_n)
\label{eq:selfen}
\ee
where $T$ is the temperature, $N$ the system size and 
$V(\vec{q},i \nu_n)$ the effective interaction resulting from a
geometric series of  bubble and ladder diagrams
\be
\label{eq:Veff}
V(\vec{q},i \nu_n) = U^2 
\bigg(\frac{3}{2} \sdw(\vec{q},i \nu_n) +
\frac{1}{2}\cdw(\vec{q},i \nu_n) -\chi(\vec{q},i \nu_n)  \bigg)
\ee
with

\bea
\label{chi}
&&\chi(\vec{q},i \nu_n)= -\frac{T}{N} \sum_{\vec{k},\omega_m}
G(\vec{k}+\vec{q}, i \omega_m +i \nu_n)
G(\vec{k}, i \omega_m) \label{eq:bubble}\\
&&\chi_{\rm c/s}(\vec{q},i \nu_n)=
\frac{\chi(\vec{q},i \nu_n)}{1\pm U\chi(\vec{q},i \nu_n)}
\eea

With respect to the FLEX approximation, the MFLEX neglects 
particle-particle
fluctuations which turn out to be 
 of minor importance 
in the parameter range we are considering, i. e.,
 close to the 
antiferromagnetic instability where the effective interaction is 
dominated by the spin-fluctuation part $\chi_{\rm s}$.\cite{PaoBickers}
This approximation is  conserving at the one-particle level\cite{BaKa},
since the self-energy is obtained as a functional derivative of a free
energy functional, containing particle-hole fluctuations only,
with respect to the Green's function. 
However, the procedure to obtain two-particle correlation functions in
a conserving way is much more complicated numerically, since functional
differentiation of the self-energy with respect to the Green's
function would include, beyond the standard ladder and RPA diagrams we are
considering, also vertex corrections\cite{bi.wh.91}. 
In the present work, we
will neglect these vertex corrections
 and thus our MFLEX
approximation is not conserving at the two-particle level.
This allows us to evaluates physical quantities at real frequencies
for larger system sizes and smaller
temperatures  allowing comparison with experimental results.
Moreover, according to Dahm and Tewordt\cite{da.te.95.1}, the corrections
coming from these additional diagrams seem to be  small in a similar
parameter range.
Since we are mainly interested in spectral densities of one- and
two-particle correlation functions at finite frequencies
we should eventually analytically continue 
Eq. (\ref{sigma}-\ref{chi}) to real frequencies by inverting 
the corresponding Laplace transformation.\cite{si.si.90}
This inversion, however, introduces 
errors for large real frequencies $\omega\gtrsim t$ due to the exponential
kernel of the  transformation.
In order to avoid these uncertainties  we will employ a recent
approach~\cite{reelleFLEX}
which deforms the frequency sums 
(Eq. \ref{sigma}-\ref{chi}) to the line $\omega+i \delta$
with $\delta \leq \pi T/2$
close to the real axis 
and carry out the self-consistent calculation directly on this line.
From this line, we can continue analytically our results to the real axis
with a much better accuracy, since the 
imaginary part of the susceptibilities
 at $\omega + i
\delta$ already shows most of the features that are present in the true
spectral functions on the real axis.

The magnetic properties of the Hubbard model  are
related in linear response to the retarded spin-spin correlation function 
\be
-\chi_{zz}(\vec q,\omega)= -i
\int_0^{\infty} d t\ e^{i\omega t}
\langle [S_z(\vec q,t)
S_z(-\vec q,0)] \rangle
\ee
with
$S_z(\vec q)=1/{2 N} \sum e^{- i \vec R_i \vec q}
(n_{i,\uparrow}-n_{i,\downarrow})$.
For simplicity, and in order to achieve larger system sizes and lower
temperatures, we will
neglect 
vertex corrections\cite{bi.wh.91}
in the coupled MFLEX equation for two-particle correlations  thus
 calculating the
spin response function with the ``bubble'' sum {\it with the dressed}
Green's functions obtained within the MFLEX formalism and use
\be
\chi_{zz}(\vec q,\omega)=\frac{2 \chi(\vec q,\omega)}{1- U \chi(\vec
q,\omega)} \; .
\label{eq:chi_zz}
\ee
Here, $\chi(\vec q,\omega)$ 
is given by Eq.~(\ref{eq:bubble})
after  continuation to the real-frequency axis. 
Neglecting these diagrams, makes the approximation not conserving, as
discussed above.

\section{Single-particle properties}
\label{single}

\begin{figure}[thb]
\centerline{\psfig{file=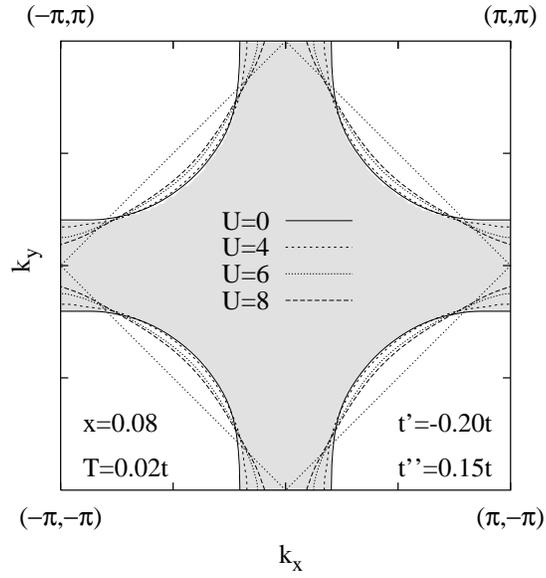,width=8cm}}
\narrowtext
\caption{\label{fig:Fermisurface_tm20p15} The Fermi-surface for different
values of $U$ for fixed temperature $T=0.02t$ and hole doping $x=0.08$.
The shaded area represents the states occupied by electrons.}
\end{figure}

\begin{figure}
\vbox{%
\centerline{\psfig{file=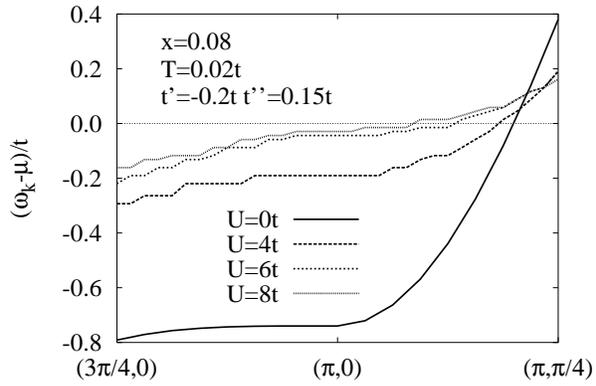,width=8cm}}
\caption{\label{fig:band_tm20p15} Quasiparticle dispersions for the 
Fermi-surfaces shown in Fig.~\ref{fig:Fermisurface_tm20p15}. 
We only show the most interesting flat part around $(\pi,0)$.
}
}
\end{figure}

We focus our study on the properties of two 
$\rm YBa_2Cu_3O_{7-\delta}$
 samples, an
underdoped one with $\delta = 0.37$ and a nearly optimally doped one
 with   $\delta = 0$.
For the sake of our comparison, however, we need to know
the appropriate value of the hole doping $x$
 to use in our Hamiltonian
 associated with these two oxygen concentrations. 
The question of how many holes go into the $\rm CuO_2$ layers for a given
oxygen content in 
$\rm YBa_2Cu_3O_{7-\delta}$
is quite controversial. 
Presland \ea~\cite{DopingParabola} 
 have suggested the 
empirical formula 
\be
T_c/T_{c, {\rm max}} = 1 - 82.6 (x-0.16)^2 
\label{eq:empiricdoping}
\ee
which relates the doping $x$ to the ratio of $T_c/T_{c, {\rm max}}$, where
$T_{c, {\rm max}}$ is the critical temperature at optimal doping.
This relation is particular useful in conjunction with measurements of the
thermoelectric power at room temperature, since this quantity shows a
generic dependence on the hole concentration.\cite{genericTEP,TEPpaper}
Thus,
a measurement of the thermoelectric power at room temperature is
uniquely related to
the doping concentration.
According to Eq.~(\ref{eq:empiricdoping}), we estimate
$x\approx0.08$ for the underdoped 
$\rm YBa_2Cu_3O_{6.63}$ sample and $x\approx0.16$
 for the fully oxygenated
$\rm  YBa_2Cu_3O_{7}$.
For these doping levels, the parameter set $t'=-0.2t$ and
$t''=0.15t$ yields a Fermi-surface in good qualitative agreement with
experiments 
in the sense that it is closed around $(\pi,\pi)$ and shows a large
curvature, 
see Fig.~\ref{fig:Fermisurface_tm20p15}. Alternative parameter sets yielding
a similar Fermi-surface are, e. g., $t'=-0.45t, t''=0$ and $t'=-0.38t,
t''=0.06t$.
Of course, for an interacting  system
one expects this  bare Fermi-surface 
to change, when the interaction
$U$ is turned on.
Indeed, in Fig.~\ref{fig:Fermisurface_tm20p15} we show the Fermi-surface for
various values of $U$, including $U=0$, obtained by our self-consistent
MFLEX
calculation at the  temperature $T=0.02 t$.
Here, the Fermi-surface is defined by the $\vec k$ points matching the
condition 
\be
\label{relfs}
\epsilon_{\vec k} + \mbox{Re }\Sigma(\vec k, \omega=0) - \mu= 0 \; .
\ee
Notice that the Fermi-surface is unambiguously defined only for $T=0$.
At finite $T$, other alternative definitions (like, e. g., the local maximum
of the spectral function at $\omega=0$, or the local maximum of
$| \nabla_{\vec k} \langle n_{\vec k} \rangle |$)
may give results differing an amount of
order $T/v_F$ from
 Eq. (\ref{relfs}).
As one can see, 
the interaction $U$ modifies the Fermi surface, as expected, especially in
the regions close to $(\pi,0)$ and 
 $(\pi/2,\pi/2)$ [and symmetrically related points]. 
By increasing the interaction from $U/t=0$ to $U/t=8$ the curvature of the 
Fermi-surface is smoothly reduced. The effect of the interaction is
thus to increase the magnetic fluctuations by pushing the Fermi-surface
closer to nesting with a wave vector 
$\vec Q$ equal or close to $(\pi,\pi)$.\cite{pr.ha.97,sh.sc.97}
On the other hand, the decrease with increasing $U$  of the area 
inside the Fermi-surface in the region close to $(0,\pi)$ 
seems to be compensated by its increase in the region close to $(\pi/2,\pi/2)$.
Whether this compensation is exact 
as suggested by the Luttinger theorem
is not clear,\cite{violation} since it is difficult to extrapolate the
result to 
$T\to 0$ were the
Fermi-surface would be well defined.
The strong effect of $U$ on the quasiparticle dispersion\cite{dispersion}
near the
point $(\pi,0)$
is shown in
Fig.~\ref{fig:band_tm20p15}. 
The main effects of the interaction
are (i) to flatten the  dispersion near $(\pi,0)$
and (ii) 
to decrease the binding energy of the quasiparticles 
$\Delta_X=E(\pi,0) -\mu$
close to  this point\cite{Putz}.
Specifically, the energy scale 
 $\Delta_X$ is seen to change
from being of the order of the bandwith $\Delta_X\approx 0.8 t$ at
$U=0$ to 
the order of the magnetic excitation 
$J=4t^2/U$ at finite $U$ (more precisely, we have $\Delta_X$ roughly
 $0.15 \cdot J$).
The effect of $U$ is thus to pin 
the quasiparticle dispersion at
$(\pi,0)$ 
to the chemical potential. This pinning is related to the onset of a pseudogap 
as was already pointed out in Refs.~\onlinecite{pr.ha.97,Altmann97}.
%%@@
Nevertheless, the flat region approaches the Fermi surface with
increasing hole density.

\section{Comparison of MFLEX with QMC}
\label{flex_qmc}

\begin{figure}
\centerline{\psfig{file=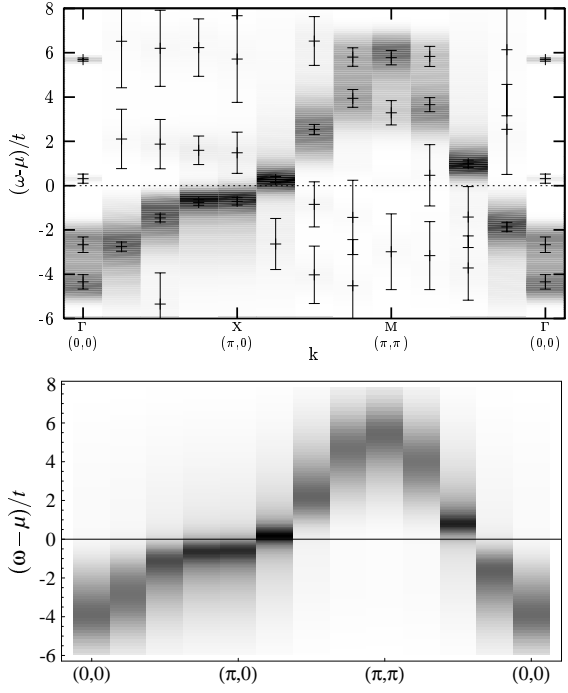,width=8cm}}
\caption{\label{fig:Akwgray} Spectral functions $A(\vec k, \omega)$
for $U=4t, t'=-0.2t,
t''=0.15t, x=0.08, T=0.33t$
as obtained by QMC (top) and MFLEX (bottom) on $8 \times
8$ systems. Dark (bright) areas correspond to a large (low)
spectral weight. Note that the energy dispersion
depends weakly on  the
temperature.}
\end{figure}

\begin{figure}
\vbox{%
\centerline{\psfig{file=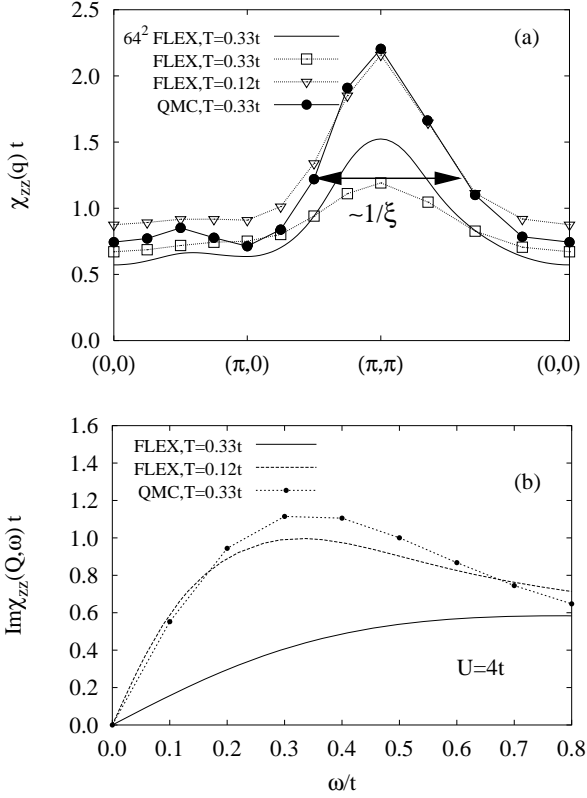,width=8cm}}%
\caption{\label{fig:FLEX-QMC-U4} (a) Static spin susceptibility
$\chi_{zz}(\vec q, \omega=0)$ along the
standard path in the Brillouin zone as obtained by MFLEX and QMC
calculations. (b) Corresponding dynamical spin susceptibility
${\rm Im}\chi_{zz}(\vec Q, \omega)$ at the antiferromagnetic momentum $\vec
Q=(\pi,\pi)$. All data are for $U=4t, t'=-0.2t, t''=0.15t$
and a $8 \times 8$ lattice
except for the MFLEX results
in (a), additionally labeled by $64^2$, which are for a $64
\times 64$ system.
}
}
\end{figure}

Before  comparing the  magnetic
properties obtained in MFLEX and QMC, we
first analyze the single-particle spectral function 
$A(\vec k, \omega)=-{\rm Im}G(\vec k,\omega+i0^+)/\pi$.
For QMC we obtain this function
by the Maximum Entropy method from the dynamical Green's function $G(\vec
k,\tau)$, while in our MFLEX approach we use the Pad\'e approximation to
continue the data from the line $\omega+i\delta$
slightly above  the real frequency axis.
Fig.~\ref{fig:Akwgray} shows the data for $U=4t$ and $T=0.33t$, where QMC is
contrasted.
This figure displays
a very good overall  agreement between both technique, thus 
strengthening our
confidence in the MFLEX approximation.
A similar comparison of the quasiparticle dispersion between MFLEX and QMC,
but for the three-band Hubbard model was carried out in Ref.~\onlinecite{Putz}.
We now turn to the magnetic properties of the Hubbard model.
Fig.~\ref{fig:FLEX-QMC-U4}(a) shows the static
magnetic susceptibility $\chi_{zz}(\vec q, \omega=0)$ 
calculated within the MFLEX approximation [Eq. (\ref{eq:chi_zz})]
compared with  QMC data for $U/t=4$ and
different system sizes along the standard path in the Brillouin zone
$(0,0)\to (\pi,0) \to (\pi,\pi) \to (0,0)$.
We first consider this quantity for the comparison since
it  does not rely on an analytical continuation to the real frequency
axis performed by the Maximum Entropy method and therefore the QMC
result  has very small
errors. 
In Fig.~\ref{fig:FLEX-QMC-U4}(a) we show the QMC results for
$U=4t, x=0.08, t'=-0.2t,
t''=0.15t, T=0.33t$ 
obtained on a $8\times 8$ lattice.
For  the same parameter set and system size, 
the corresponding MFLEX results
(squares) show a smaller susceptibility and a  much smaller
peak at $(\pi,\pi)$. This means that the antiferromagnetic fluctuations
are  underestimated within
the MFLEX approximation  compared with those in the
QMC simulations. 
The question arises whether this  is due to the fact that we have
neglected vertex corrections
 or, rather, to the approximation itself.
Dahm \ea~\cite{DahmVertex}
determined the leading contributions 
to 
the vertex corrections
 (up to $O(U^2)$) and found that
these tend to {\em reduce} 
the spin susceptibility. 
On the other hand, the comparison of our results with QMC
 apparently shows that  the
spin-spin correlation function calculated {\it without}
vertex corrections
 are 
 {\em underestimated} with respect to the (in principle exact) QMC  results.
 Thus, either the quality of the perturbative results
worsens when including vertex corrections
 or higher order contributions to
these diagrams become important at our intermediate values of $U$. %\\
In the case of a non self-consistent RPA calculation\cite{BulutUeff} 
it was shown that the introduction of an
effective $U^*<U$  restored a good comparison with QMC calculation.
This is because in the RPA approximation {\em without} self-consistent
Green's function magnetic fluctuations are {\em overestimated} by the RPA
denominator. 
In our  calculations, 
$\chi$ is renormalized  self-consistently by the  
renormalization of $G$, which, in turn, reduces 
$\chi$ especially near its maximum at $(\pi,\pi)$.
However, this reduction  overcompensates for the enhancement
due to the RPA denominator and this
is the reason why 
 magnetic fluctuations are underestimated in this self-consistent
 calculation.
Taking example from Ref. \onlinecite{BulutUeff},
one could  think to introduce an effective $U^*$ {\em greater}
than $U$ in order to compensate for the reduction of magnetic
fluctuations. This procedure would also be a simplified version of 
the 
``pseudopotential approach'' introduced by Bickers et al.\cite{bi.wh.91}
to include  parquet diagrams in an effective way.
However, 
we have verified that
increasing $U$ from $4t$ to $8t$ at a fixed high temperature $T=0.33t$ 
results in a slight {\it decrease} of $\chi_{zz}(\vec q)$. The opposite
occurs at lower temperatures like $T=0.05t$.
 Thus 
the introduction of a temperature-independent 
$U^*$ (which is needed to extrapolate QMC results to lower
temperatures) cannot improve the results for $\chi$. %\\

On the other hand, since magnetic
fluctuations are very sensitive to the temperature, we 
introduce an {\em effective temperature $T^\star<T$} and compare our MFLEX
results calculated with the temperature $T^\star$ with QMC results
calculated at the temperature $T$.
The physical motivation for this ansatz is that, due the closeness of
the system to the
Mott-Hubbard transition at half-filling,  antiferromagnetic fluctuations
are very strong\cite{pr.ha.95}. 
These fluctuations, while fully captured by QMC,
 are underestimated by the MFLEX approximation which
is not able to describe the metal-insulator transition appropriately.
 Since  antiferromagnetic
fluctuations are very sensitive to the temperature, the shortcoming of the
MFLEX approach can be removed by a reduction of the temperature. 
Indeed, our comparison with
QMC results for $U=4t$ is greatly improved if one takes a
scale factor $R$ such that $T^\star=T/R$ with $R=2.5$. 
The MFLEX results at this temperature $T^\star$
can be seen in
Fig.~\ref{fig:FLEX-QMC-U4}(a) (triangles) to compare quite well
with QMC data at temperature $T$. This is especially true for the
correlation length as indicated by the arrow.
From the same figure, one can also infer the importance 
of using a method which
allows  to increase the system size.
Here indeed, we also
present the MFLEX results for a $64 \times 64$ 
system at $T=0.33t$. For this increased system 
the static susceptibility
shows significantly more pronounced magnetic
fluctuations than the  $8\times 8$ results with the same $T$.
It is also interesting that the
$64\times 64$ MFLEX results resolve the small peak between $(0,0)$ and 
$(\pi,0)$ observable in QMC although the magnetic response is still smaller
than the one obtained with QMC at the same temperature.

\begin{figure}[thb]
\centerline{\psfig{file=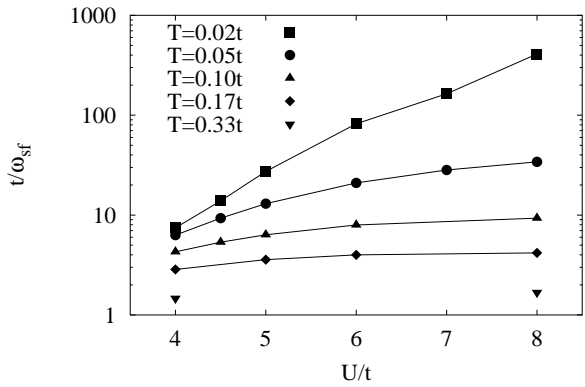,width=8cm}}%
\caption{\label{fig:wsf_vs_U} The {\em inverse} of the magnetic energy scale
$\wsf$ as a function of $U$ for various temperatures ($x=0.08, t'=-0.2,
t''=0.15$). A SDW instability manifests itself 
 by $\wsf \to 0$.}
\end{figure}

The same renormalized temperature $T^\star$ can be used in order to
achieve a good agreement  with QMC of the imaginary part of the {\it dynamic} 
spin susceptibility
$\mbox{Im}\chi_{zz}(\vec Q,\omega)$. 
In Fig.~\ref{fig:FLEX-QMC-U4}(b) we show 
$\mbox{Im}\chi_{zz}(\vec Q,\omega)$ for 
$U=4t$   obtained
with MFLEX and QMC on the same small $8\times 8$ systems.  
As for 
 the static susceptibility,
the MFLEX approximation yields a much
smaller value for
$\mbox{Im}\chi_{zz}(\vec Q,\omega)$
than the QMC result, whenever calculated with the same temperature
 $T$. 
However, if the MFLEX results are calculated at
 the reduced temperature $T^\star$ introduced above 
the agreement is drastically improved. In particular,
 the slope of 
$\mbox{Im}\chi_{zz}(\vec Q,\omega)$
at $\omega=0$ (related to $T_1$, see below)
as well as the  position $\wsf$ of the maximum
 agree very well.
Thus, renormalizing the temperature in the MFLEX calculations leads to a
considerable improvement of the perturbative results as compared with
 QMC.
We have verified that  the effective temperature $T^\star$ is 
related to the true temperature $T$ by approximately the same scale
factor $R$ also at higher temperature. For $T=t$ we obtain an
optimized agreement with QMC results similar to
Fig.~\ref{fig:FLEX-QMC-U4} with $R\approx 2$. This gives us confidence
that the renormalization factor will be appropriate to describe
{\it qualitatively} the spin correlation function also for smaller
temperatures at least within an  error of $\sim 20\%$
which we anyway allow for the fit to experiments. 
Finally, the temperature renormalization does not affect the agreement
of the quasiparticle dispersion as shown in
Fig.~\ref{fig:Akwgray}, since the latter depends only
weakly on the temperature.
The importance of the temperature for the spin fluctuations becomes clear in
Fig.~\ref{fig:wsf_vs_U}, where we show  the (logarithm of) the 
inverse of the frequency $\wsf$ where 
$\mbox{Im}\chi_{zz}(\vec Q,\omega)$ is maximum,
 as a function of $U$ for different $T$.
$1/\wsf$ is
 an indication of the strength of the antiferromagnetic fluctuations,
 since it diverges in the spin-density-wave state.
\noindent 
For high temperatures, $1/\wsf$ is only weakly $U$-dependent  while for
very low temperatures $\wsf$ vanishes exponentially with increasing $U$
indicating the SDW instability (at $T=0$) for
            $U\to \infty$.\cite{DahmVertex} 
Below, we will show that our results with $U=4t$, 
(whereby the MFLEX temperature is set to the renormalized temperature
$T^\star=T/\RT$), 
agree quite
 well with experimental results on $T_1$ and
reasonably well with
$T_{2G}$ at low temperatures.

\begin{figure}[thb]
\vbox{%
\centerline{\psfig{file=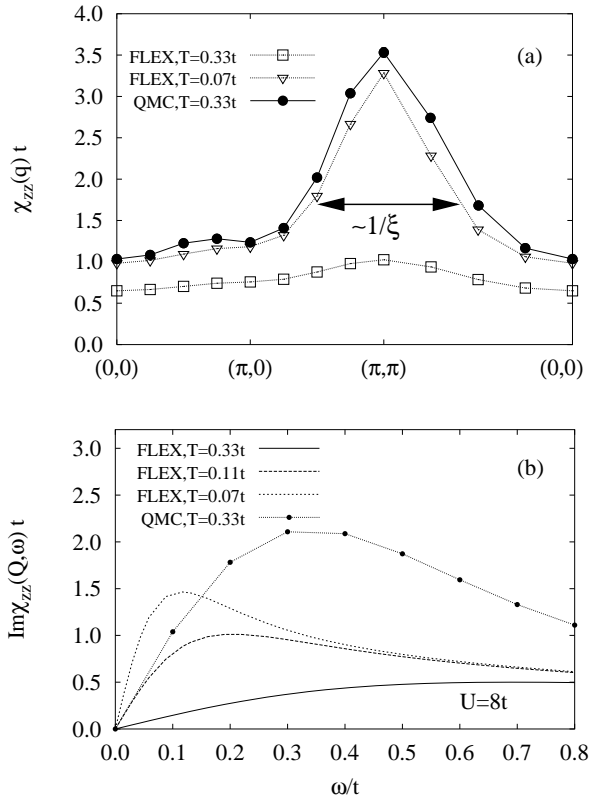,width=8cm}}%
\caption{\label{fig:FLEX-QMC-U8} Same as Fig.~\ref{fig:FLEX-QMC-U4}, but
for $U=8t$ and $x=0.10$.%
(a) Static spin susceptibility 
(b) Corresponding dynamical spin susceptibility
}
}
\end{figure}

Nevertheless,
 it is believed that 
the properties of the cuprate materials at and close to half filling 
are better described by a larger $U/t$\cite{pr.ha.95}.
We will thus also show the results for
 $U/t=8$. 
However, applying perturbation theory  when the interaction is
as large as the bandwidth ($U = 8t \approx  W$) is questionable. 
Nevertheless, it is tempting again to
compare the diagrammatic results with QMC and thus use the 
MFLEX calculation to
extrapolate the QMC data to lower temperatures and larger system
sizes.
The static susceptibility  for $U=8t$ on  $8\times 8$ clusters 
is presented in Fig.~\ref{fig:FLEX-QMC-U8}(a).
For $T=0.33t$, $\chi_{zz}(\vec q)$ obtained in the present approximation is rather
flat and structureless and does not compare well with QMC results.
We thus use again our strategy of renormalizing the temperature by a
factor $\RT$, which must necessarily
 depend on $U$ and is expected to 
be larger for increasing $U$. 
As one can see from  Fig.~\ref{fig:FLEX-QMC-U8}(a), we need a
temperature renormalization factor $\RT$ of about $5$ in order to have a
good agreement for
the static spin susceptibility 
$\chi_{zz}(\vec q)$. On the other hand,
from  Fig.~\ref{fig:FLEX-QMC-U8}(b) one can
  see that it is not possible
to find an appropriate temperature renormalization which makes
the imaginary part (i.e. the dynamical properties) 
${\rm Im}\chi_{zz}(\vec Q,\omega)$ to agree 
with
the QMC result. If one requires that only the slope at $\omega=0$ coincides
(which is the important quantity necessary to calculate $T_1$) we need
$\RT\approx 3$.
This value of $\RT$ does not coincide with the one obtained for the
static correlation function. 
This clearly shows the difficulty of using this diagrammatic approach
for such large interaction strength.
For comparison with experiments in Sec. \ref{flex_exp}, we will 
use an {\em intermediate} temperature renormalization factor $\RT=4.$

\section{Comparison with  experiments}
\label{flex_exp}

Most of the available experimental results on 
the magnetic properties of the high-$T_c$ materials 
are extracted from
nuclear magnetic resonance (NMR)
 and 
inelastic neutron scattering (INS) 
  studies. While INS measures directly
the $\vec q$ and $\omega$-resolved ${\rm Im}\chi_{zz}(\vec q, \omega)$,
NMR,  as a local probe, determines weighted averages of the susceptibility
$\chi_{zz}(\vec q, \omega)$ for  $\omega\to0$  over the
whole Brillouin zone: 
Specifically, 
the spin-lattice relaxation time $T_1$ probes the 
inverse of the slope
of ${\rm Im}\chi_{zz}(\vec q, \omega)$
 for $\omega \to 0$ and the spin-echo decay time
$T_{2G}$ the  inverse of the static susceptibility.
NMR and INS experiments on LSCO and YBCO have revealed a lot of remarkable
properties: 
($i$) strong antiferromagnetic fluctuations persisting
in the normal as well as in the
superconducting phase up to the optimally doped regime,  
($ii$) a suppression of 
${\rm Im}\chi_{zz}(\vec q, \omega)$ at small $\omega$ 
attributed to a spin gap
opening at low temperatures in metallic YBCO, 
and 
($iii$) a sharp resonance peak
at 41~meV and $\vec q = \vec Q$
for optimally doped YBCO below $T_c$.\cite{Rossat91,Fong97}
The spin gap manifests itself in INS
measurements with a depression of the magnetic response
at low energies and low
temperatures. 
NMR measurements  agree with this spin gap and show
 a depression of $T_1 T$ below $T^*_{\rm INS}$ which is about $T_c$ for the
overdoped and larger than $T_c$ for the underdoped
samples.\cite{Horvatic93}

To relate the spin-lattice relaxation time $T_1$ to the spin
susceptibility $\chi_{zz}$, we adopt the approach by
Shastry, Mila and Rice~\cite{SMR}  describing the hyperfine coupling of the
$\rm Cu^{2+}$ spins with the different nuclei in  the unit cell, which
leads 
to the expression:\cite{ZBP}

\be
\frac{1}{T_1T}= \lim_{\omega \to 0} \frac{1}{2N} \frac{k_B}{\hbar}
\sum_{\vec q} F_c(\vec q) \frac{{\rm Im}\chi_{zz}(\vec q, \omega)}{\hbar
\omega}
\label{eq:T1T}
\ee
\noindent
where the form factor $F_c(\vec q)$ results from the Fourier transform 
of the hyperfine interaction
\be
F_c(\vec q)=\left\{A_{ab} + 2 B [\cos q_x + \cos q_y]\right\}^2 
\; .
\label{eq:formfactor}
\ee
Here, we consider the case where the applied static magnetic field is
perpendicular to the $\rm CuO_2$ planes. A different geometry of the
experiment would require a different form factor.\cite{BP}

The transverse relaxation rate $T_{2G}$  is related 
with the static spin susceptibility
through the Gaussian
component of the spin echo, as pointed
out independently
by Thelen and Pines~\cite{ThelenPines} and Takigawa~\cite{Takigawa94}:

\bea
T_{2G}^{-2}&=&\frac{0.69}{128 \hbar^2}
\Bigg[ \frac{1}{N} \sum_{\vec q} F_{\rm eff}^2(\vec q) 
\chi^2_{zz}(\vec q,0) \nonumber \\
&&-\bigg(\frac {1}{N}
\sum_{\vec q} F_{\rm eff}(\vec q)\chi_{zz}(\vec q,0)\bigg)^2\Bigg]
\label{eq:T2G}
\eea

with a different
form factor $F_{\rm eff}(\vec q)$ for $T_{2G}$,  obtained form
Eq.~(\ref{eq:formfactor}) by replacing $A_{ab}$ with $A_{c}$. 

The unknown hyperfine coupling constants $A_{ab}, A_{c}, B$
are extracted from Knight shift
experiments.
Here, we adopt the values recently given in the analysis by Barzykin and 
Pines~\cite{BP} and set
$A_{ab}=0.84 B, A_c=-4B$ and finally the energy scale 
$B=3.82\times 10^{-7} {\rm eV}$. These are similar to values given by other
authors~\cite{ZBP,BulutPRL90,MMP}
Note, that both relaxation times give complementary information: while
$T_1$ probes the slope of the imaginary part of $\chi_{zz}(\vec q,\omega)$
for  $\omega\to 0$, 
$T_{2G}$ depends on the static susceptibility $\chi_{zz}(\vec q,\omega=0)$. 
Since NMR probes the local environments of the spins, all momenta $\vec q$
contribute in principle to the relaxation rates as can be seen from
Eqs. (\ref{eq:T1T},\ref{eq:T2G}).
However, in the presence of a large antiferromagnetic correlation
length, the $q$ points close to the AF point $(\pi,\pi)$ will give the
largest contribution to these expressions.

In the following analysis,
we choose $U=4.5t$ which turns out to give the best agreement 
with the experimental results regarding magnetic properties. Moreover,
as discussed in Sec.\ref{flex_qmc}, 
for intermediate $U$ ($U\lesssim 6t$) only  is it possible
to have a good comparison with QMC results with a unique
temperature-renormalization factor $\RT$. For larger $U$ ($U\gtrsim 7t$)
this cannot be made unambiguously.
The energy scale  $t$ is fixed by 
taking $t=250$ meV, so that the bandwidth is
$\sim 2 {\rm eV}$ as observed in photoemission experiments.
Furthermore, we take the same temperature renormalization factor
$\RT=2.5$ as found for $U=4t$, since $\RT$ does not change much from $U=4t$
to $U=4.5t$.
The experimental results for $T_1$ and $T_{2G}$ of the two $\YBCO$
 samples are taken from Imai et al.~\cite{im.sl.93} for
$\rm YBa_2Cu_3O_{7}$  and from Takigawa~\cite{taki.94} for
$\rm YBa_2Cu_3O_{6.63}$ 
(these data  are collected in Ref.~\onlinecite{BP}). 

\begin{figure}[thb]
\vbox{
\centerline{\psfig{file=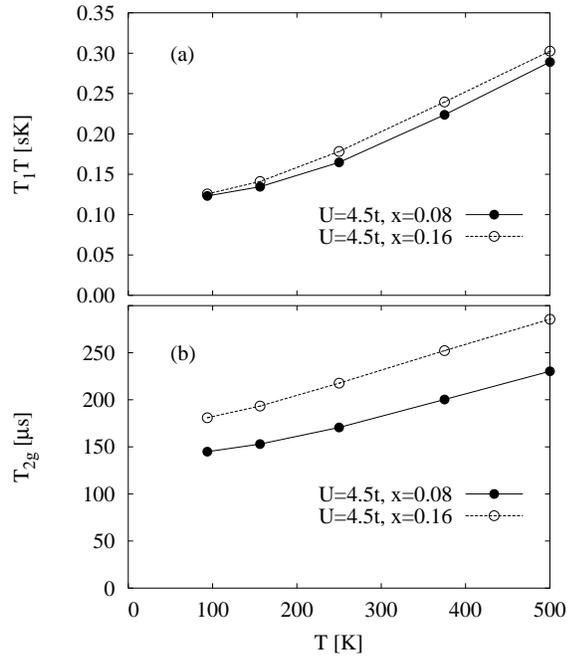,width=8cm}}
\caption{\label{fig:T1T} (a) $T_1T$ and (b) $T_{2G}$ 
NMR relaxation times calculated in the MFLEX
approximation for $U/t=4.5$, $t'/t=-0.2$, and $t''/t=0.15$, 
and two different doping levels, representing 
$\rm YBa_2Cu_3O_{6.63}$ ($x=0.08$) and $\rm YBa_2Cu_3O_{7}$
($x=0.16$).
}}
\end{figure}

In Fig.~\ref{fig:T1T}(a), we show the spin-lattice
relaxation time $T_1$ multiplied by $T$ as obtained from  the
MFLEX calculations on a $64 \times 64$ lattice for $x=0.08$ and $x=0.16$
corresponding to the two samples $\rm YBa_2Cu_3O_{6.63}$ and $\rm
YBa_2Cu_3O_{7}$, respectively. 
The figure shows that 
$T_1T$ is proportional to $T$ for high temperatures, 
while for lower temperatures 
it tends to a constant. 
While the linear behavior at high temperatures agrees with
experiments, for $T$ smaller than a certain value, $T_1 T$ 
 should increase again due to the occurrence of a spin gap.
In our calculation we are not able to see this gap behavior, possibly
because we cannot reach the temperature where the gap sets in or
because of the limitations of our approximation.
However, a precursor of the spin gap is seen in the flattening of
$T_1T$ 
at low temperatures.
The linear behavior of $T_1T$
with temperature (indicating $T_1 = {\rm const.}$)
away from the spin gap regime is actually very well
reproduced by MFLEX calculations in a wide range of 
parameters.\cite{BulutPRL90,DahmVertex}
Aiming at carrying out a {\it semi-quantitative} comparison with
experiment, it is thus natural
to fit the two parameters of the linear behavior, namely, the $T\to0$
extrapolation and the slope of $T_1 T$ versus $T$.
The theoretical and experimental values obtained for these two
parameters 
are listed in Tab.~\ref{table}.
Notice that we took into account the temperature renormalization
factor $\RT$ and modified the slope  accordingly.
In agreement with the experimental data, 
we find that the extrapolated $T_1T$ is only slightly  larger for the 
$\rm YBa_2Cu_3O_{7}$ sample than for the underdoped
$\rm YBa_2Cu_3O_{6.63}$ sample.
Moreover, the bigger slope present in the overdoped
sample  suggests  
that the two functions cross at  some higher temperature $T_{\rm
 cross}$ ($\approx 50 K$),
as observed in
experimental results.

The  $\rm ^{63}Cu$ spin-echo decay time $T_{2G}$ 
calculated according to Eq. (\ref{eq:T2G})
is shown in 
Fig.~\ref{fig:T1T}(b).
Again, the measured 
data show  approximately
a linear-$T$ behavior
in the range between $\sim$100K and 300K, in
agreement with our theoretical results. 
For a semi-quantitative comparison 
with experiments
we again extract the slope and the $T\to0$
extrapolation and show the results in Tab.~\ref{table}.
Note, that increasing the hole doping results in a shift 
of $T_{2G}$
to larger values, while the slope remains almost the same,  in agreement with
the experimental findings.

While for $T_1T$ both the slope and the $T\to 0$-extrapolated values agree
quite well 
with the experiments (within $25\%$) 
in the case of $T_{2G}$,
the agreement is good only for 
 the slope, while the $T\to 0$ extrapolated values is too large,
 especially for the underdoped sample. In principle, we could try to
 adjust this extrapolated value by decreasing $|t'|$ but this would
 worsen the results for $T_1T$.
To show that a deviation of $25\%$ is a good result, 
   we consider the effect of a small change in $t'$.
We thus 
  include the data for $U=4.5t,
t'=-0.25t, t''=0.15t$ in Tab.~\ref{table} showing that a
change of 25\% in $t'$ results
in more than 100\% changes in the  $T\to0$ extrapolated values of $T_1T$ and
$T_{2G}$. Notice that the slopes are not very much affected by such a change.
By increasing $U$ or decreasing $|t'|$ the  values of $T_1T$ and
$T_{2G}$ extrapolate to smaller and eventually to  negative 
values. This signals that the system approaches a SDW instability.
In Tab.~\ref{table} we also include results for $U=8t$, but with $t'=-0.4,
t''=0.15$
and a temperature renormalization factor of $\RT=4$ as discussed in
Sec.~\ref{flex_qmc}.
The results are worse than the $U=4.5t$ ones, especially concerning the
$T\to 0$ extrapolation. Notice that the latter are quite sensitive to
$t'$ and could be improved by increasing $|t'|$. On the other hand,
the slope is essentially independent on $t'$ and cannot be improved in
the same way.

Notice that the too large value for the $U=4.5t$, 
$T\to 0$-extrapolated $T_{2G}$ 
is due to the fact that this quantity is extremely sensitive to the
value of $U$. For example, for $U=4t$ one would have obtained
$T_{2G}$ $\sim 200 \mu s$ for  $T \to 0$.
A fine tuning of $U$ could  fix the
extrapolated value 
of $T_{2G}$ more accurately, although this would put $T_1T$ off.

Although the  value $U/t=4.5$ seems to give the best agreement with the
magnetic properties at finite doping, 
the same value of $U/t$
 does not 
 reproduce correctly
the insulating behavior for the half filled model. Electron energy loss
and optical experiments have revealed a charge transfer gap of
$\sim 1.7 {\rm eV}$ for YBCO \cite{FinkRomberg}
which would require rather large values for $U/t\geq 8$ for our chosen
value of $t=250 {\rm meV}$.
Moreover, INS experiments on the antiferromagnetic parent compound 
$\rm YBa_2Cu_3O_6$ \cite{Rossat91} showed that this system is well
described by a spin-1/2 antiferromagnetic Heisenberg model with an
exchange coupling of $\sim 0.125 {\rm eV}$.
Using $t=250 {\rm meV}$ and $J=4 t^2/U$ one needs a value of
$U/t\approx 8$.
On the other hand, our diagrammatic approach cannot be well reliable
for such a large value of $U/t$ as discussed in Sec.~\ref{flex_qmc},
even when using a temperature renormalization factor extracted from
the comparison with QMC data at high temperatures.
For this reason, we cannot rule out that a more appropriate
calculation could give a good comparison with experiments on $T_1T$ and
$T_{2G}$ also for $U/t\approx 8$.
Another possibility could be that
the effective $U/t$ at finite doping may be reduced
with respect to the one at half filling due to the screening of the
doped carriers.

\begin{figure}[thb]
\centerline{\psfig{file=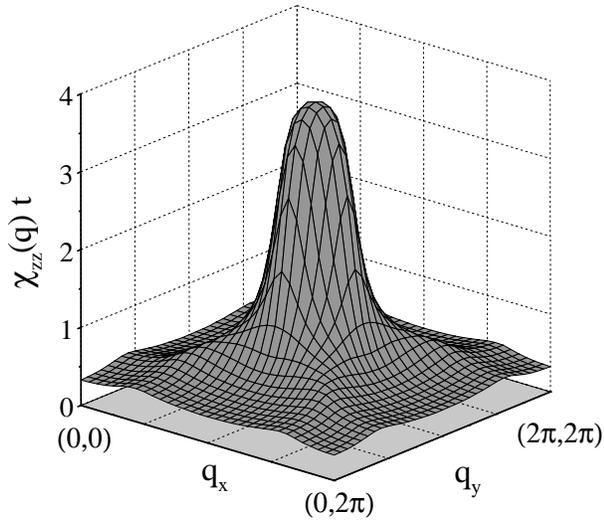,width=8cm}}
\caption{\label{fig:chi3d} Static susceptibility in the Brillouin zone
for $U=4.5t,x=0.08$ and $T=150{\rm K}$.Although this function is strongly
peaked at $\vec Q =(\pi,\pi)$ indicating a commensurate spin response,
it is better fitted by four incommensurate peaks.}
\end{figure}

\begin{figure}[thb]
\centerline{\psfig{file=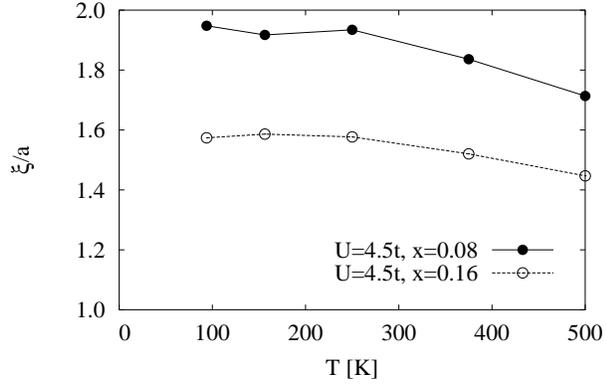,width=8cm}}
\caption{\label{fig:corrlength} Correlation length obtained by fitting
the static susceptibility along the line from
$(\pi,0)$ to $(\pi,\pi)$ with a 
superposition of 4 incommensurate Lorentzian-shaped peaks symmetrically
located around $\vec Q$.}
\end{figure}

We now consider the correlation length of these systems.
This  can be inferred
 from the $\vec q$-dependence of the
static spin susceptibility $\chi_{zz}(\vec q,\omega=0)$,
plotted
in Fig.~\ref{fig:chi3d}. 
for 
$U=4.5t, t'=-0.2, t''=0.15, x=0.08, T=150K$.
Since $\chi_{zz}(\vec q,\omega=0)$ is strongly peaked at the
antiferromagnetic wave vector $\vec Q = (\pi,\pi)$,
the susceptibility appears to be
commensurate. 
This is in contrast with LSCO which  clearly shows maxima 
at the incommensurate points 
$Q'\equiv (\pi +q_o',\pi)$ (and symmetric points).\cite{Mason92,in.prb}
Experimentally,
it is not clear whether an incommensurability is seen in  
 the spin response of the
YBCO materials. Early INS
experiments~\cite{Rossat91} and some more recent ones~\cite{Bourges96}
suggest a
commensurate structure, while other authors
report experimental data that are better
fitted by a superposition of 
Lorentzian curves peaked at the 
 four 
equivalent incommensurate points  $(\pi, \pi\pm
q_o)$ and $(\pi\pm q_o, \pi)$ ~\cite{Tranquada} or 
at $(\pi\pm q_o, \pi\pm q_o)$,\cite{Dai97}
where the
incommensurability $q_o$ is material and doping dependent.
To extract the correlation length $\xi$ and the incommensurability
$q_o$ of the calculated 
spin-spin correlation
function we model $\chi_{zz}(\vec q,\omega=0)$  by the superposition of 
four Lorentzian curves with width $1/\xi$ 
peaked at the points $(\pi, \pi\pm q_o)$ and  $(\pi\pm q_o, \pi)$.
We carry out the fit along 
the line  $(\pi,q)$ with $0\leq q \leq\pi$
where the  model function discussed above reads

\bea
&&\chi_{zz}(q,\omega=0)=\frac{\Gamma}{(1/\xi)^2+(q-Q+q_o)^2}
\nonumber \\
&&+ \frac{\Gamma}{(1/\xi)^2+(q-Q-q_o)^2}+
\frac{2\Gamma}{(1/\xi)^2+(q-Q)^2+q_o^2}
\eea

with $Q = \pi$.
The results for the correlation length
are shown in Fig.~\ref{fig:corrlength}. As observed in
experiments, the correlation length $\xi$ is of the order of $1-2$
lattice spacings\cite{Rossat91,Tranquada,Bourges96} 
 and temperature independent for
$T\lesssim250{\rm K}$. It decreases for increasing doping levels.
Although 
the single maximum in the curve of 
Fig.~\ref{fig:chi3d} suggests a commensurate 
structure, it is better fitted  with an
incommensurability $q_o \approx 0.5$. 
This agrees with recent INS experiments, where  an
incommensurability of $q_o \approx 0.4$ was  suggested 
to better reproduce the spin susceptibility in 
$\rm YBa_2Cu_3O_{6.6}$.\cite{Dai97.2}

If we define $\xi$ to be just the half width half maximum (HWHM), the
correlation length is even smaller ($\xi/a \approx 1$ for $x=0.08$) but
still  temperature independent.
Our finding of a relatively small correlation length $\xi/a \sim 1--2$ 
 is in contrast with
the phenomenological NAFL treatment of the NMR relaxation times by 
Barzykin and Pines,\cite{BP}  where rather large correlation lengths of about
$7a$ for $\rm YBa_2Cu_3O_{6.63}$ and $2a$ for $\rm YBa_2Cu_3O_{7}$
were necessary for satisfying fits of the experiments.
That these correlation lengths are too large in comparison with experiments
was already pointed out in a later critical reexamination by Zha,
Barzykin and Pines.\cite{ZBP}
A relatively small correlation length is also the reason why the
temperature dependence of $\xi$ and $T_{2G}$ are different.

\begin{figure}[thb]
\centerline{\psfig{file=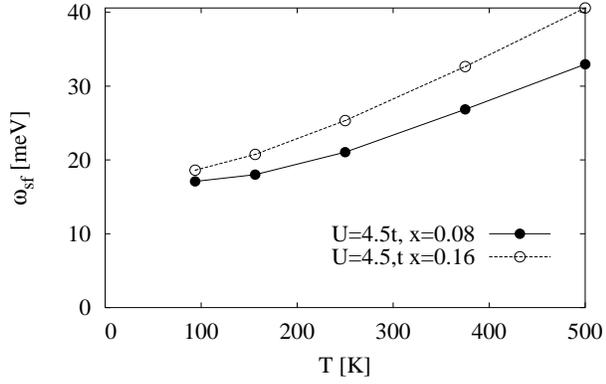,width=8cm}}
\caption{\label{fig:wsf} The magnetic energy scale $\wsf$ defined to be the
frequency where the dynamical susceptibility at $\vec Q$ is maximal.}
\end{figure}

Finally,  in Fig.~\ref{fig:wsf} we plot as a function of $T$
the magnetic energy scale $\wsf$ defined to be the energy
where ${\rm Im}\chi_{zz}(\vec q, \omega)$ takes its maximum.
The calculated values are between 20 and 40~meV, in good agreement with the
energy of the magnetic resonance peak found in
experiments.\cite{Fong97}
A comparison between Fig.~\ref{fig:T1T} and 
Fig.~\ref{fig:wsf} suggests that 
$T_1T$ and $\wsf$ show the same linear
temperature dependence, in agreement with the NAFL theory.
For $T \to 0$,  $\wsf$ also tends to a constant 
which decreases with decreasing $|t'|$ or increasing $U$, i. e.,
approaching 
the SDW instability.

\section{Summary and Conclusions}
\label{summary}

In summary, we have studied the electronic and 
magnetic properties of an underdoped and an overdoped $\YBCO$ sample 
with $\delta=0.37$
and $\delta=0$, respectively. We started from the two-dimensional 
Hubbard model including longer-ranged hopping processes 
to describe the correlation effects in these materials. 
Since it is essential 
to reach both low temperatures  and a fine spectral resolution
for a qualitative 
 comparison with experiments, we employed the
MFLEX approximation, which amounts to 
neglecting particle-particle fluctuations
and vertex corrections in
the FLEX approximation.
We checked the quality of this approach by  comparing with
 Quantum-Monte Carlo results at higher temperatures. 
We found that for not too large Hubbard interactions $U$ ($U \lesssim W/2$)
the agreement
between the MFLEX and QMC results can be
 considerably improved by introducing a
$U$-dependent
renormalized temperature $T^{\star}=T/\RT$.
We then use this temperature 
renormalization factor $\RT$ in order to extrapolate
the QMC results towards the temperature necessary for an analysis of the
experiments. 
One should be aware of the fact that this approach
is uncontrolled and may be ineffective, since
the temperature renormalization factor 
$\RT$ found at the temperatures accessible by QMC ($T/t\sim 0.33$) may
not hold for lower temperatures and thus the extrapolation may fail.
On the other hand, as we already mentioned, 
the temperatures accessible to QMC 
are far too high and thus 
an  extrapolation scheme to lower temperatures is mandatory.

In the search for appropriate parameters of the Hubbard model to
describe  qualitatively and semi-quantitatively 
the magnetic properties of $\YBCO$,
we find a 
fair agreement with experimental results on $T_1$ and $T_{2G}$, 
(within an average error of less than $ 20\%$)
for the parameter set
$U=4.5 t, t'=-0.2t, t''=0.15t$ and $t=250 {\rm meV}$. For this value
of $U/t$, we need  a temperature
renormalization $R\approx 2.5$, as inferred from the comparison with QMC. 
Our calculations  describe in a qualitative
way also  the
shape of the Fermi-surface, and the 
flat quasiparticle energy dispersion 
near $(\pi,0)$ which  approaches the Fermi surface at
optimal doping \cite{ma.de.96}.
Moreover, we have a reasonable description of
(i) the slope and $T\to 0$-extrapolated values of $T_1T$,
where $T_1$ is the spin lattice relaxation time, (ii) the slope of the
spin-echo decay time $T_{2G}$ vs. temperature, (iii) the  
correlation length $\xi\approx$ 1--2 lattice spacings, 
(iv) the $q$-dependence of the  spin
response, which appears commensurate due to a single maximum at 
$\vec Q=(\pi,\pi)$, but which  
is better fitted by a superposition of four Lorentzian curves peaked
at  incommensurate
peaks, and (v) the typical size of the
magnetic excitation energy scale of 20--40~meV. 
Finally, our calculation is not inconsistent with
 a parameter
set with stronger coupling,  like $U/t=8$,
as seems to be the case for the cuprate materials,
 provided one increases 
the value of the next-nearest-neighbor hopping.
However, our procedure is less reliable in this parameter regime.

Our results thus indicate the importance of introducing
finite values for longer-range hoppings 
$t'/t$ and $t''/t$  for the sake of a qualitative description
of {\it magnetic} properties. This is in agreement 
 with  numerical calculations on the $t-J$ model
which show the relevance  
of $t'/t$ and $t''/t$ for an appropriate description
of {\it single-particle} 
properties.\cite{to.ma.94,an.li.95,be.ch.96,xi.wh.96,ed.oh.97.d,ki.wh.98}.
Notwithstanding the importance of longer-range hoppings appearent from
these  works,
 it is difficult to establish whether other parameters, like
possibly an interplane coupling or longer-range interactions may 
be physically 
more important and thus describe the experiments more appropriately
even without introducing $t'$ and $t''$.

\acknowledgments
We thank J.~Schmalian for many useful discussions and 
together with  M.~Langer, S.~Grabowski, and
K.~H.~Bennemann
for providing us with their
code for the real-frequency approach to the MFLEX approximation
described in Ref. \onlinecite{reelleFLEX}.
Financial support by the Bavarian High-$T_c$ Program FORSUPRA (GH and WH) 
and by the EC TMR Project N. ERBFMBICT950048 (EA) 
is 
acknowledged. We finally thank the 
Lebniz-Rechenzentrum in Munich, the ZAM in J\"ulich and the HLRS in
Stuttgart for providing us with CPU time for the numerical
calculations.

\vbox{%
\widetext%
\begin{table}
\begin{tabular}{|l|r|r|r|r|}
&\multicolumn{2}{c|}{$T_1T$}& \multicolumn{2}{c|}{$T_{2G}$}\\
\tableline
& Slope & $T\to 0 $ & Slope & $T\to 0$ \\ 
&  [s]&  [sK] & [$\mu \rm sK^{-1}$]& [$\mu$s]\\ \tableline
$\rm YBa_2Cu_3O_{7}$&$ 5.0\cdot 10^{-4}$&
$7.5 \cdot 10^{-2}$ & 0.18 & 84\\
MFLEX $x=0.16$&$ 4.0 \cdot 10^{-4}$& $7.9 \cdot 10^{-2}$ & 0.26 & 153\\
\tableline
$\rm YBa_2Cu_3O_{6.63}$& $4.0 \cdot 10^{-4}$& 
$8.5 \cdot 10^{-2}$ & 0.14 & 23\\
MFLEX $x=0.08$& $3.2 \cdot 10^{-4}$& $8.4 \cdot 10^{-2}$ &0.19 & 124\\
\tableline
MFLEX $x=0.08$& $1.8 \cdot 10^{-4}$& $2.2 \cdot 10^{-1}$ &0.11 & 198\\
$U=4.5t, t'=-0.25t, t''=0.15t$&&&&\\
\tableline
MFLEX $x=0.08$& $2.8 \cdot 10^{-4}$& $1.9 \cdot 10^{-2}$ &0.24 & 102\\
$U=8t, t'=-0.40t, t''=0.15t$&&&&

\end{tabular}
\caption{\label{table} Slope and extrapolated value 
for $T$ greater than the spin gap obtained by
fitting a straight line to the measured\protect\cite{im.sl.93,taki.94}
(labelled with 
$\rm YBa_2Cu_3O_{7-\delta}$)
and calculated (MFLEX) data for $T_1T$ and $T_{2G}$ with $U/t=4.5, t'/t=-0.2, t''/t=0.15$. 
Comparison with the results for $U/t=4.5, t'/t=-0.25, t''/t=0.15$
and $U/t=8, t'/t=-0.4, t''/t=0.15$ is shown.}

\end{table}
}

%%\bibliography{myrefs,preprints,mypublications,localbiblio}  %%c%%

\begin{thebibliography}{10}
\bibitem{lspacing}{
We measure the momenta in units of the lattice spacing.}%LCIT
\bibitem{dispersion}{
We define the band dispersion $\omega(\vec k)$ by the maximum in $\omega$ of
  the spectral function $A(\vec k, \omega)$ for each $\vec k$.}%LCIT
\end{thebibliography}
%%\bibliographystyle{myprsty} %%c%% 

\def\nonformale#1{#1}
\def\formale#1{}
\def\spa{} \def\spb{}
\spa

\end{document}